\documentclass[%
 aip,
 jmp,%
 amsmath,amssymb,amsfonts,amsthm,
 reprint,%
]{revtex4-1}
\usepackage{slashed}
\usepackage{graphicx}
\usepackage{dcolumn}
\usepackage{bm}
\usepackage{chngcntr}

\counterwithin*{equation}{section}
\nocite{*}

\begin{document}


\title[A Statistical Analysis of Anharmonic Gases]{A Statistical Analysis of Anharmonic Gases}

\author{Nikhil Kalyanapuram}%
 \email{nikhilkaly@cmi.ac.in}
\affiliation{ 
Chennai Mathematical Institute, H1, SIPCOT, Siruseri, Kelambakkam, Tamil Nadu.
}%

\date{\today}

\begin{abstract}
In this paper, we consider an anharmonic perturbation to the harmonic oscillator in the classical and the quantum regimes. We analyse a relativistic particle subjected to such a potential and then proceed to study a gas of such particles. In the first case, the partition function is computed and stated. In both cases, expressions for the average energy are obtained.
\end{abstract}

\maketitle

\begin{quotation}

\end{quotation}

\section{Introduction}
The equation of state of a normal gas is obtained by considering non-relativistic free particles. Further refinements of such considerations are obtained by studying the case of free particles subjected to a harmonic perturbation or treating them in the relativistic regime. The partition function for a free particle under a harmonic perturbation in three dimensions is simple, we simply cite it here for reference,

$$
Z_{1} = \frac{V}{(2\pi\hbar)^3}\left(\frac{2\pi k_B T}{\omega\sqrt{m}}\right)^3.
$$

Here, $m$ and $\omega$ are used to denote the characteristic mass and frequency of the particle considered. 

For the case of a relativistic particle under a harmonic perturbation of amplitude $A$, the following integral must be evaluated,

\begin{widetext}
$$
Z_2 = \frac{VQ}{(2\pi \hbar)^6}\int_{-\infty} ^\infty \mathrm{d}^3 p\mathrm{d}^3 \xi \exp\left(-\frac{c}{k_B T}\sqrt{m^2c^2 + p^2} - \frac{m\omega^2}{2k_B T}|\xi|^2\right).
$$
\end{widetext}

The computation is done in the appendix, giving,

\begin{widetext}
\begin{equation}
Z_2 = 4\pi VQ\left(\frac{ c}{(2\pi \hbar)^2}\sqrt{\frac{2\pi mk_B T}{ \omega^2}}\right)^3\left(\frac{k_B T}{mc^2}K_{0}\left(\frac{mc^2}{k_B T}\right) + 2\left(\frac{k_B T}{mc^2}\right)^2 K_1\left(\frac{mc^2}{k_B T}\right)\right)
\end{equation}
\end{widetext}

where,

$$
Q = \frac{4}{3}\pi\left(\left(\frac{m\omega^2A^2}{2c^2}\right)^2 - m^2c^2\right)^{3/2}.
$$

$Q$ is the volume of a sphere in momentum space whose radius is given by the maximum momentum attainable by the particle by virtue of it's vibratory motion. Note that since $A$ may be arbitrarily large, $Q$ is formally infinite.

Here, no error is suffered by the replacement of $A$ by $\infty$ in the integral as the amplitude may be chosen to be arbitrarily large, keeping in mind that $Q$ will drop out of all physically relevant quantities.

Using the well known expressions for the thermodynamic variables as obtained from their partition functions, one may easily obtain the desired equations of state. 

That said, we may consider the situation when such a particle is subject to an external potential which leads to an anharmonic potential disturbing it's motion. That supplies the motivation for the next section.

Thereafter, we consider the quantum mechanical effects derived from such a disturbance to second order in perturbation theory. The relevant partition function is then obtained and studied in context. Standard expressions for the frequency distribution of blackbody radiation are generalised and treated appropriately. 

Finally, applications of the developed theory are explored, particularly in the context of cosmological distributions of background radiation. Anisotropies are studied using the relevant formulae. 

Throughout the paper, new equations derived are stated with minimal detail as to the prior computations involved. The more non-trivial of these are detailed in the appendix.

\section{The Classical Anharmonic Oscillator}
In this section, we study the classical anharmonic oscillator and some of it's statistical properties. In view of this goal, we consider the following system. The system is composed of a single free particle localized in a defined volume $V$, with translational and vibrational degrees of freedom. The Hamiltonian for such a system is supplied by,

\begin{equation}
    H = H_{kin} + H_{vib}.
\end{equation}

Here, $H_{kin}$  is a purely momentum dependent quantity. In our model, $H_{vib}$ however takes the following form,

\begin{equation}
    H_{vib} = \frac{m\omega^2}{2}r^2 + \lambda r^4.
\end{equation}

For a minima to exist obviously, we must have $\lambda>0$. Now, we focus on that part of the partition function that is determined by the vibrational modes. Employing the symbol $Z_{\xi,1}$ for this component, we obtain,

$$
Z_{\xi,1} = 4\pi\int_{0}^{\infty}r^2 \exp(-\frac{m\omega^2}{2k_B T}r^2 - \frac{\lambda}{k_B T} r^4)\mathrm{d}r .
$$

We obtain as the solution,

\begin{equation}
     Z_{\xi,1} = 4\pi \left(\frac{k_B T}{\lambda}\right)^{3/4} F\left(\sqrt{\frac{m^2 \omega^4}{64 k_B T \lambda}}\right)
\end{equation}

where the function $F$, calculated in the appendix, is given by,

\begin{widetext}
\begin{equation}
    F(x) = - e^{2x^2}\left(\frac{1}{4\sqrt{x}}K_{\frac{1}{4}}(2x^2) + 2x\sqrt{x}K_{\frac{1}{4}}(2x^2) + 2x\sqrt{x}K'_{\frac{1}{4}}(2x^2)\right).
\end{equation}
\end{widetext}

Defining one more function,

\begin{equation}
    G(x) = xK_{0}\left(x\right) + 2x^2 K_1\left(x\right)
\end{equation}

the partition function for a relativistic particle under an anharmonic perturbation will be given by the formula,

\begin{equation}
    a_{0}F\left(\sqrt{\frac{m^2 \omega^4}{64 k_B T \lambda}}\right)G\left(\frac{mc^2}{k_B T}\right).
\end{equation}

where,

$$
a_0 = 16\pi^2 \frac{(mc)^3 V V_P }{(2\pi \hbar)^6}\left(\frac{k_B T}{\lambda}\right)^{3/4}.
$$

Here, $V_P$ is the entire momentum space volume, as the vibrational momentum now admits of values over all of momentum space. It is again formally infinite, but drops out of physically meaningful answers.

With the formal calculation complete, we are now in a position to compute the equation of state and study the analytic properties of the above partition function. 

As a first step, we may study the average energy of such a particle. Happily, taking the logarithm and effecting the derivative eliminates the unphysical volumes, leaving us with the expression for the average energy as,

\begin{widetext}
\begin{equation}
    \overline{E} = \frac{3k_B T}{4} - \sqrt{\frac{m^2\omega ^ 4 k_B T}{256 \lambda}}\frac{F'\left(\sqrt{\frac{m^2 \omega^4}{64 k_B T \lambda}}\right)}{F\left(\sqrt{\frac{m^2 \omega^4}{64 k_B T \lambda}}\right)} - mc^2 \frac{G'\left(\frac{mc^2}{k_B T}\right)}{G\left(\frac{mc^2}{k_B T}\right)}.
\end{equation}
\end{widetext}

Having completed this short calculation, we now move on to the main concern of this letter, namely the study of the consequences generated by the introduction of an anharmonic perturbation to a quantum gas.

\section{The Quantum Mechanical Anharmonic Oscillator}
As a first step, let it be noted that the introduction of an anharmonic perturbation to a harmonic oscillator may be realised in nonlinear perturbations to quantum fields. In service of this possibility, we investigate the cases of massless and massive gases, with the goal of eventually seeking applicability in the field theoretic context.

Less abstractly, consider the following Hamiltonian,

\begin{equation}
    H = \frac{p^2}{2m} + \frac{m\omega^2 x^2}{2} + \mu x^3 + \lambda x^4 
\end{equation}

where $m$ is some characteristic mass scale, $\mu$ and $\lambda$ are coupling constants of small order.

The unperturbed Hamiltonian admits of a diagonalisation that supplies the following energy eigenvalues,

$$
E = n\hbar\omega + \frac{\hbar\omega}{2}.
$$

Applying time-independent perturbation supplied the following second order perturbations to the energy levels, the computational details of which are relegated to the appendix.

\begin{equation}
    \delta E_{1} = -\frac{\mu^2\hbar^2}{16m^3\omega^4}\left(n^2 + 6n + 5\right)
\end{equation}

\begin{equation}
    \delta E_{2} = \frac{3\lambda\hbar^2}{4m^2\omega^2}\left(2n^2 + 2n + 1\right)
\end{equation}

Now, we are left with the following energy dependence on the number of excitations,

\begin{equation}
    E_n = A(\omega)n^2 + B(\omega)n + C(\omega)
\end{equation}

where the coefficients are meromorphic functions of $\omega$ given by,

\begin{equation}
    A(\omega) = -\frac{\mu^2\hbar^2}{16m^3\omega^4} +  \frac{3\lambda\hbar^2}{2m^2\omega^2} 
\end{equation}

\begin{equation}
    B(\omega) = -\frac{3\mu^2\hbar^2}{8m^3\omega^4} +  \frac{3\lambda\hbar^2}{2m^2\omega^2}+ \hbar\omega 
\end{equation}

\begin{equation}
    C(\omega) = -\frac{5\mu^2\hbar^2}{16m^3\omega^4} +  \frac{3\lambda\hbar^2}{4m^2\omega^2} + \frac{\hbar\omega}{2}.
\end{equation}

With the foregoing expressions at our disposal, we shall proceed to compute the grand canonical partition function for the normal classes of relativistic gases. The first case to be considered is that of a massless Bose-Einstein gas. Treating each quantum as an anharmonic oscillator subject to the spectrum (3.3), we first note the following sum,

$$
Z^{BE}_{\omega,0} = \sum_{n=0}^{\infty} \exp\left(-\beta A(\omega)n^2-\beta B(\omega)n -\beta C(\omega)\right).
$$

where $\beta = \frac{1}{k_B T}$. The convergence of this sum is easily established by noting that,

$$
|Z^{BE}_{\omega,0}| \leq e^{-\beta C(\omega)}\sum_{n=0}^{\infty}\exp\left(-\beta B(\omega)n\right).
$$

which is satisfied if $A(\omega)$ is positive. The convergence of the above sum is ensured only by the satisfaction of $B(\omega)>0$. 

Physically, this corresponds to the following argument. If the coefficients given above are allowed to assume negative values, then given a sufficiently large number of particles, the energy values will assume negative values. In light of this, the positivity conditions are implemented in the integration itself using Heaviside functions.

Employing the standard definition for the average energy, the following expression is obtained,

\begin{widetext}
$$
\overline{E} = \frac{V}{\pi^2 c^2}\int_{0}^{\infty}\omega^2\frac{ A(\omega)n^2+ B(\omega)n + C(\omega)}{Z^{BE}_{\omega,0}}\Theta\left(\beta A(\omega)\right)\Theta\left(\beta B(\omega)\right)\mathrm{d} \omega.
$$
\end{widetext}
Define the following dimensionless variable,

$$
y = \frac{\hbar\omega}{k_B T}
$$

and the characteristic functions,
\begin{widetext}
$$
f_n(y) = \frac{a_A(n^2 + 6n)}{y^4} + \frac{a_B (2n^2+2n)}{y^2} + ny
$$
\end{widetext}

where,

$$
a_A = -\frac{\hbar^6 \mu^2}{16m^3 k_B^5 T^5}
$$
and
$$
a_B = \frac{3\lambda\hbar^4}{4m^2k_B^3T^3}.
$$

Note that the zero point energies arising from quantum fluctuations about the vacuum have been neglected. 

We now have,

$$
\beta A(\omega) = \frac{a_A}{y^4} + \frac{a_B}{y^2} = g_2(y)
$$

$$
\beta B(\omega) = \frac{6a_A}{y^4} + \frac{2a_B}{y^2}= g_1(y).
$$

Now, doing away with the zero point contributions in the energy density one obtains,

\begin{widetext}
$$
\mathcal{E}_0 = \frac{k_B^4 T^4}{\hbar^3 \pi^2 c^3}\int^\infty_0\frac{\sum_{n=0}^{\infty}f_{n}(y)\exp(-f_n(y))}{\sum_{n=0}^{\infty}\exp(-f_n(y))}\Theta(g_1(y))\Theta(g_2(y))y^2 dy.
$$

For a massive Bose-Einstein gas we obtain the following expression,

$$
\mathcal{E}_m = \frac{mc^2 k_B T}{2\hbar^3 \pi^2 c^3}\int^\infty_0\frac{\sum_{n=0}^{\infty}f_{n}(y)\exp{(-f_n(y))}}{\sum_{n=0}^{\infty}\exp(-f_n(y))}\Theta(g_1(y))\Theta(g_2(y))y\left(\sqrt{\left(\frac{k_B T}{Mc^2}\right)^2 y - 1}\right) dy.
$$

The integration over the theta functions may be readily evaluated, giving the two simple expressions,

$$
\mathcal{E}_0 = \frac{k_B^4 T^4}{\hbar^3 \pi^2 c^3}\int^\infty_{3\kappa^2}\frac{\sum_{n=0}^{\infty}f_{n}(y)\exp(f_n(y))}{\sum_{n=0}^{\infty}\exp(f_n(y))}y^2 dy.
$$

$$
\mathcal{E}_M = \frac{Mc^2 k_B T}{2\hbar^3 \pi^2 c^3}\int^\infty_{3\kappa^2}\frac{\sum_{n=0}^{\infty}f_{n}(y)\exp{(-f_n(y))}}{\sum_{n=0}^{\infty}\exp(-f_n(y))}y\left(\sqrt{\left(\frac{k_B T}{Mc^2}\right)^2 y - 1}\right) dy.
$$
\end{widetext}

where, $\kappa^2 = \frac{-a_A}{a_B}$.

One point may be made regarding this minimum threshold. Treated as a function of the lower limit of integration, one notes that the partition functions diverge below $3\kappa^2$. 

One must note the arbitrariness of the characteristic constants $m$, $\mu$ and $\lambda$. We may note that the combination $\frac{\mu}{\lambda m}$ has units of frequency squared. In effect, a bound on this may be secured by establishing an empirical infrared cut-off.

The variation of the partition function as these constants are varied supplies a continuous motion in the space of available theories. We will not pursue this direction here, but it was felt that the point was worth mention.

We will now proceed to evaluate the energy density in the case of the massless Bose-Einstein gas in a specified limit. The limit is that which permits the neglect of higher order contributions in the denominator, which essentially allows for the replacement of the denominator by the following expression,

$$
1-e^{-y}
$$

The numerator admits of the following expansion,

\begin{widetext}
\begin{equation}
   \sum_{n=0}^{\infty} \left(\frac{a_A(n^2 + 6n)}{y^4} + \frac{a_B (2n^2+2n)}{y^2} + ny\right)\exp\left(-\frac{a_A(n^2 + 6n)}{y^4} - \frac{a_B (2n^2+2n)}{y^2} - ny\right)
\end{equation}
\end{widetext}

Expanding out the exponentials with negative powers of $y$ and explicitly writing out the summations gives us the following series,

\begin{widetext}
\begin{equation}
   \sum_{n=0}^{\infty}\sum_{i=0}^{\infty}\sum_{j=0}^{\infty}\frac{1}{i!}\frac{1}{j!}(-1)^{i+j}(a_A(n^2+6n)^i(a_B(2n^2 + 2n))^j \left(\frac{a_A(n^2 + 6n)}{y^{2 + 4i + 2j}} + \frac{a_B (2n^2+2n)}{y^{4i + 2j}} + ny^{3 - 4i - 2j}\right)\exp\left( - ny\right)
\end{equation}
\end{widetext}

Firstly, we define the following functions,

\begin{widetext}
$$
\begin{aligned}
F_{i,j, n} = & a_A(n^2 + 6n)n^{1+4i+2j}(3n\kappa^2)^{-1-2i-j}e^{-\frac{3}{2}(n+1)\kappa^2}W_{-(1+2i+j), -\frac{1+4i+2j}{2}}(3n\kappa^2) +\\
&a_B (2n^2+2n)n^{-1+4i+2j}(3n\kappa^2)^{-2i-j}e^{-\frac{3}{2}(n+1)\kappa^2}W_{-(2i+j), -\frac{-1+4i+2j}{2}}(3n\kappa^2)+ \\
&n^{-4+4j+2j}n(3n\kappa^2)^{-\frac{-3+4i+2j}{2}}e^{-\frac{3}{2}(n+1)\kappa^2}W_{-\frac{-3+4i+2j}{2}, -\frac{-4+4i+2j}{2}}(3n\kappa^2)
\end{aligned}
$$

$$
\begin{aligned}
F_{i,j, n} = & a_A(n^2 + 6n)(n+1)^{1+4i+2j}(3n\kappa^2)^{-1-2i-j}e^{-\frac{3}{2}(n+1)\kappa^2}W_{-(1+2i+j), -\frac{1+4i+2j}{2}}(3(n+1)\kappa^2) +\\
&a_B (2n^2+2n)(n+1)^{-1+4i+2j}(3(n+1)\kappa^2)^{-2i-j}e^{-\frac{3}{2}(n+1)\kappa^2}W_{-(2i+j), -\frac{-1+4i+2j}{2}}(3(n+1)\kappa^2)+ \\
&(n+1)^{-4+4j+2j}n(3(n+1)\kappa^2)^{-\frac{-3+4i+2j}{2}}e^{-\frac{3}{2}(n+1)\kappa^2}W_{-\frac{-3+4i+2j}{2}, -\frac{-4+4i+2j}{2}}(3(n+1)\kappa^2)
\end{aligned}.
$$

Employing Ref \cite{grad} equation 3.381 (6), we obtain the desired formal solution ,

\begin{equation}
    \mathcal{E}_{0} = \frac{k_B^4 T^4}{\hbar^3 \pi^2 c^3}\sum_{n=0}^{\infty}\sum_{i=0}^{\infty}\sum_{j=0}^{\infty}\frac{1}{i!}\frac{1}{j!}(-1)^{i+j}(a_A(n^2+6n)^i(a_B(2n^2 + 2n))^j \left(F_{i,j,n} - G_{i,j,n}\right)
\end{equation}
\end{widetext}

Now, $a_A$ and $a_B$ play the role of dimensionless coupling constants, determining a class of theories. They are not independent however, and depend on the infrared cut off $\kappa^2$. It may be noted that in the limit where these go to zero, the sum collapses leaving behind the standard expression for blackbody radiation. This may be inferred directly from the expansion (3.8).

\section{Concluding Remarks}
The foregoing considerations are mostly formal, and applications are yet to be considered. As a speculative note, there are perhaps applications to be had in determining corrections to blackbody radiation in nonlinear electrodynamics.

\section*{Acknowledgements}

\section*{Appendix A}
In this appendix, we consider the evaluation of (1.1). To do so, note that the substitution $p = mc\sinh(x)$ may be effected, leaving one with the integral,

$$
\int_{0}^{\infty}\sinh^2(x)\cosh(x) \exp(-z\cosh(x))\mathrm{d}x
$$

with $z=\frac{mc^2}{k_B T}$ and ignoring prefactors.

The integral is then evaluated in terms of modified Bessel functions using,

$$
\int_{0}^{\infty}\sinh^2(x)\exp(-z\cosh(x))\mathrm{d}x = K_1(z)/z.
$$

The prefactor $Q$ is only formally defined. The expression given may be obtained by noting the length of the momentum vector corresponding to maximum potential energy,

$$
\sqrt{p^2 + m^2 c^2} = \frac{m\omega^2 A^2}{2c}.
$$

The volume of the corresponding sphere in momentum space yields the desired expression.

Now we consider the evaluation of the integral leading to (2.3). TO do so, we consider the fiducial integral,

$$
\int_0^\infty \exp(-x^4 - 4a x^2)\mathrm{d}x
$$

Formally differentiating with respect to $-4a$ will yield the desired integral, provided suitable definitions $x^4 = \frac{\lambda}{k_B T}r$ and $a  = \sqrt{m^2 \omega^4}{64k_B T \lambda}$ are made. The final answer then follows by application of 3.323 (3) from Ref \cite{grad}.

Finally, we consider the evaluation of the corrections to the energy eigenvalues. This is done using standard time independent perturbation theory. The formulae used are,

$$
\Delta E^{(1)}_n = \left(\psi_n , H_I, \psi_n\right)
$$

$$
\Delta E^{(2)}_n = \sum_{k\neq n}\frac{|\left(\psi_n , H_I, \psi_k\right)|^2}{E^{(0)}_n - E^{(0)}_k}
$$

where $H_I$ is the interaction part of the total Hamiltonian. The standard calculation then supplied the energy corrections (3.2) and (3.3), using the second order and first order formulae respectively.
\begin{widetext}

\end{widetext}

\end{document}